\documentclass[11pt,oneside,a4paper]{article}
\usepackage[left=1in,right=1in,top=1in,bottom=1in]{geometry}
\usepackage{amsmath,amssymb}

\usepackage{hyperref}

\usepackage{graphics}
\usepackage{graphicx}
\usepackage{caption}
\usepackage{subcaption}

\usepackage{setspace}
\usepackage{float}

\hypersetup{colorlinks=true, linkcolor=blue, citecolor=red, bookmarks=false, pdfstartview={FitH}}

\begin{document}

\title{\bf Using Maple + GRTensorII in teaching basics of General Relativity and Cosmology}

\author{
{\bf Ciprian A. Sporea\thanks{E-mail:~ciprian.sporea89@e-uvt.ro} , Dumitru N. Vulcanov}\thanks{E-mail:~vulcan@physics.uvt.ro}\\
\ \ \\
{\small West University of Timi\c soara, Faculty of Physics,}\\
{\small V. Parvan Ave. no. 4, 300223, Timi\c soara, Romania} \\
}

\date{\small \today}

\maketitle

\begin{abstract}
In this article we propose some Maple procedures, for teaching purposes, to study the basics of General Relativity (GR) and Cosmology. After presenting some features of GRTensorII, a package specially built to deal with GR, we give two examples of how one can use these  procedures. In the first example we build the Schwarzschild solution of Einstein equations, while in the second one we study some simple cosmological models.
\ \ \\


{\bf PACS (2010):} 98.80.-k $\bullet$ 04.30.-w $\bullet$ 01.40.Ha \\

\end{abstract}

\pagenumbering{roman}

\tableofcontents

\newpage

\pagenumbering{arabic}

\section{Introduction}

Cosmology (i.e. the modern theory of Universe dynamics) has became in the last two decades an attractive field of human knowledge including also an intense media campaign. This was possible, among other reasons, also because in this time period several space missions performed cosmological measurements -like COBE, WMAP, Plank, BICEP2- thus transforming cosmology from a pure theoretical field also into an experimental one. Thus teaching cosmology, even at undergraduate level, in physics faculties comes to be a compulsory  topic. Cosmology is based on two major pillars \cite{8},\cite{13}: astrophysics as a phenomenological tool and general relativity (GR) as the main theory, thus making it, unfortunately, difficult to teach to undergrad students. The mathematical structure or GR is based on differential geometry \cite{9}, \cite{10} and learning it means that the student must first get familiar with the main instruments of differential geometry (such as tensor calculus on curved manifolds, Riemannian curvature, metric connection, covariant derivative, etc), which often means cumbersome and lengthy hand calculations. To illustrate these facts let us remind that GR is based on field equations known as Einstein equations, namely
\begin{equation}\label{EE}
R_{\mu\nu}-\frac{1}{2}g_{\mu\nu}R+ \Lambda g_{\mu\nu}=-\kappa T_{\mu\nu}
\end{equation}
where $R_{\mu \nu}$ is the Ricci tensor, $g_{\mu\nu}$ the metric tensor and $T_{\mu\nu}$ represents the stress-energy tensor. We denoted by $\Lambda$ the cosmological constant and $\kappa=8\pi G/c^4$. The components of the Ricci tensor are given by \cite{8},\cite{9}
\begin{equation}\label{Ricci}
R_{\mu\nu}=\partial_\lambda\Gamma^\lambda_{\ \mu\nu}-\partial_\nu\Gamma^\lambda_{\ \mu\lambda}+\Gamma^\lambda_{\ \mu\nu}\Gamma^\sigma_{\ \lambda\sigma}-\Gamma^\sigma_{\ \mu\lambda}\Gamma^\lambda_{\ \nu\sigma}
\end{equation}
where  $\Gamma^\lambda_{\ \mu\nu}$ are the so called Chrisstoffell symbols wich in Riemannian geometry \cite{10} describe the structure of the curved space-time underlying GR and the associated metric tensor is compatible with the connection described by the above Chrisstoffel symbols, namely
\begin{equation}\label{Chris}
\Gamma^\lambda_{\ \mu\nu} = g^{\lambda\sigma}\left( \partial_\mu g_{\nu\sigma} + \partial_\nu g_{\mu\sigma} - \partial_\sigma g_{\mu\nu}\right)
\end{equation}
All these make even the most determined students to lose interest in studying cosmology (and GR too). Dozens of pages with hundreds
of terms containing partial differentials to be hand processed could scare any student (and not only!).

As today students are by passing day more skilled in computer manipulation and with a more and more advanced practice in programming,  the modern teaching of GR and cosmology should use intensively computer facilities for algebraic programming, tensor manipulation and of course numerical and graphical facilities \cite{11}, \cite{14}, \cite{15}. The use of computer algebra was in the view of physicists even since the beginning of computer science both for teaching and research purposes. Computer algebra (or algebraic programming codes) evolved from early days of REDUCE package (see for example \cite{19},\cite{20}, \cite{11}) till recent developments using integrated platforms as Maple and Mathematica in different fields of physics, not only in general relativity (see for example \cite{11}, \cite{17}, \cite{18} and more recently \cite{22}, \cite{21}).

Some years ago \cite{4} we published our experience in this using the REDUCE platform. Unfortunately, in the last years REDUCE lost the market in favour of more intergraded and visual platforms such as Maple \cite{1} and Mathematica, thus we adapted our experience and program packages to Maple and we've made use of the free package GTTensorII \cite{2} adapted for doing GR.

The aim of the present article is to report our new experience in this direction. The article is organised as follows. The next section introduces the main features of  GRTensorII in doing tensorial symbolic computation in GR and Riemannian geometry. Section  no. 3 describes the way we can obtain an exact solution of Einstein equations. We used again, as in the main classical text on GR, the Schwarzschild solution. This is the most famous solution used today intensively in describing the motion in the solar system and for studying black-holes physics \cite{12}. The last section is dedicated to describe how one can use Maple and GRTensorII for cosmology (and teaching it). In both these two above sections we gave the main Maple commands which can be put together to have short programs to be used during the computer lab hours and even during the lectures. The article ends with a short section where we present the main conclusions and some ideas for future developments.

\section{Short presentation of the GRTensorII package }

GRTensorII is a computer algebra package built within the Maple platform as a special set of libraries \cite{1}. It is a free distributed package (see \cite{2}) and it is adapted for dealing with computer algebra manipulation in general relativity. Thus it is  designed for dealing with tensors and other geometric objects specific to Riemannian geometry (a metric that is compatible with the connection, symmetric connection - torsion free manifolds). In what follows, we will present some of the main features offered by GRTensorII.
The library is based on a series of special commands  all starting with  $"gr"$ (for example $grcalc, grdisplay, gralter, grdefine$, etc) for dealing with a series of (pre)defined geometric objects such as the metric tensor, Ricci tensor and scalar, Einstein tensor, Chrisstoffell symboles, etc.

To start the GRTensorII package one must type in a (new) Maple session the following commands
\begin{verbatim}
> restart; > grtw();
\end{verbatim}
The restart command causes the Maple kernel to clear its internal memory so that Maple acts (almost) as if just started. The second command initializes the GRTensorII package and gives some information about the current version.

The GRTensorII library allows us to build our own space-time manifold. The most easiest way to do this is by creating a metric tensor
$ g_{\mu\nu} $ with the help of $>makeg( )$ command. The result of operating this command will be a special ASCII file containing the main information about the metric and stored in a folder called "metrics" within the GRTensorII library. The folder "metrics" contains also a collection of predefined metric files distributed with the package (for egz. schw.mpl, schmidt.mpl, vdust.mpl, etc).

Another way in which one can specify a space-time manifold is by loading a predefined metric from the "metrics" folder. This can be done with the help of two commands
\begin{verbatim}
> qload(metricName);
> grload(metricName, meticFile);
\end{verbatim}
The geometry built using $>makeg()$, $>qload() $ and $>grload()$ commands will fix the  background on which all the later operations and calculations will be performed.

One of the main advantages of GRTensorII library is that it allows us to do complicated operations on tensorial objects, regardless of how
many indices those objects poses. The main command that permit us to do those calculations is

$>grcalc( objectSeq )$ that calculates the components of tensors;

For example $>grcalc ( R(dn,dn), R(up,dn,dn,dn) )$ ask the program to calculate the covariant components of the Ricci tensor $ R_{\mu\nu} $ and the components of the standard curvature Riemann tensor $ R^{\sigma}_{\mu\nu\lambda} $.

The command $ >grdisplay()$ can be used to display the components of GRTensorII objects which have been previously calculated for a particular space-time. Before displaying the calculated components of an object it is indicated to use the command $>gralter()$ in order to simplify them. The  $>grcalc()$ calculates all the components of a given object, so if one wants to calculate only a specific component then it can use the command $>grcalc1 ( object, indexList )$. For example: $>gecalc1 ( R(dn,dn,dn,dn), [t, r, \theta, \phi])$.

Besides the predefined objects that exist in GRTensorII we can also define new objects (scalars, vectors, tensors) with the help of the command $>grdef()$. For example

$>grdef ( `G2\{a\ \ b\} := R\{a\ \ b\} - (1/2)*Ricciscalar*g\{a\ \ b\} + Lambda*g\{a\ \ b\}`)$

defines a contravariant two index tensor, $G2_{a b}$, which it is explicitly assigned to an expression involving a number of previously defined (or predefined) tensors. The syntax in $> grdef()$ command follows naturally the usual tensorial operations which defines the new objects.

Another important command of GRTensorII is $> grcoponent()$ which allows us to extract a certain component of a tensorial object. The extracted component can be used as a standard Maple object for later processing (symbolically, graphically and numerically).

Although GRTensorII was designed initially for Riemannian differential geometry  it can be easily extended to other types of geometries, such as ones with torsion or higher order alternative theories of gravity \cite{5}, \cite{6}.

We end this section by making the important observation that using GRTensorII does not impose any restriction in using all the numerical, graphical and symbolic computation facilities of Maple (as it happens with other packages even for Maple). Thus we can combine al these facilities for an efficient use of the Maple platform.

\section{Example 1: Schwarzschild type solutions}

General relativity and its applications (such as cosmology) are based on Einstein equations (\ref{EE}) as main field
equations. They have  many exact solutions, although these second order nonlinear differential equations have no unique analytical
general  solution. The most famous exact solution of Einstein equations is the  Schwarzschild solution \cite{8}, \cite{9} describing the gravitational field around a pointlike mass M (or outside a sphere of mass M). This solution is used today for describing the black-hole dynamics and was used in the first attempts in applying GR to  the motion of planets and planetoids in our solar system.

It is obvious that from a pedagogical point of view finding an exact solution of Einstein equations  could be a good introductory lesson in applications of GR. Next we will derive this solution following the natural steps :

- identifying the symmetries of the system for which we build the solution;

- building a metric tensor compatible with the above symmetry;

- building the shape of the stress-energy tensor components (if any exists);

- calculating the Ricci tensor and the components of Einstein equations;

- solving the above equations after a close inspection of them.

These above steps could be done manually and usually it takes several hours of hard calculations (even straightforward and even for an experienced person). Our advise for anyone who wants to teach GR and/or cosmology is to do this traditional step with the students. It will be  a good  lesson and a motivation to proceed in using algebraic computing facilities (here Maple+GRTensorII).

Thus the above  steps are clearly transposable in computer commands in Maple+ GRTensorII.  For the Schwarzschild solution the symmetry is clearly  spherical  and static (no time dependence including the time inversion, namely $t  \rightarrow  -t$ ). Thus we will use a spherical symmetric  metric tensor as \cite{8}:
\begin{equation}\label{sferic}
ds^2=e^{2\lambda(r)}dt^2 + e^{2\mu(r)}dr^2 + r^2\left( d\theta^2 + sin^2(\theta)d\phi^2\right)
\end{equation}
in spherical coordinates  $(t,r,\theta,\phi)$, where $\lambda(r)$ and $\mu(r)$ are the two unknown functions of the radial coordinate $r$ to be found  at the end.  Thus  the student already in front of a computer or station having started a Maple session will be guided to compose the next sequence of commands
\begin{verbatim}
> restart; grtw();
> makeg(sferic);
>......
> grdisplay(metric); grdisplay(ds);
\end{verbatim}
where  after the two  commands  for starting the GRTensortII the command $> makeg$ will create the ASCII file $sferic.mpl$ containing
the information on the metric we will built. The series of dots above represent those steps  where the user has to answer with the type of the metric, symmetry and of course its components one by one.   The last two lines given above are calculating the metric and displays its shape in the form of a matrix.  After this we can continue to do some calculations or to close the session. The metric we produce can be loaded anytime later in another sessions.

The next step of our demonstrative program will be to point out and calculate the Einstein equations, but not before introducing the stress-energy components. It is obvious that in this case the strass-energy components are cancelled as we calculate the gravity field outside the source (the pointlike mass or a sphere). Thus in this case we will solve  the so called vacuum Einstein equations i.e. $R_{\mu \nu}=0$ \cite{9}.

In  this view we will built a sequence of Maple commands (for a new session):
\begin{verbatim}
> restart; grtw();
> qload(sferic); grcalc(R(dn,dn));
> gralter(R(dn,dn),simplify); grdisplay(R(dn,dn));
\end{verbatim}
where after loading the metric tensor with the command $>qload$ we calculate the Ricci tensor components, simplify and display them
 as a 4x4  matrix (using the last $>grdisplay$ command).

It is a good and interesting experience, before proceeding with the solving of Einstein eqs. to insert in the above lines the next ones
\begin{verbatim}
> grcalc(Chr(up,dn,dn)); grdisplay(Chr(up,dn,dn));
\end{verbatim}
immediately after loading the metric tensor with $> qload$ which very fast calculates and displays the 40 components of the Chrisstoffel symbols using relation (\ref{Chris}).

Of course if we want maximum effect of these, before calculating with Maple and GRTensorII we advise the teacher to calculate, by hand, together with the students at least 3 of 4 of these components. This will take some time of hard calculations and many mistakes when done for the first time.

To continue it is now much more simple to extract the components of the Ricci tensor one by one as Maple objects in order to process them to solve the obtained equations. It will be a sequence of $>grcomponent$ commands, namely :
\begin{verbatim}
> ec0:=grcomponent(R(dn,dn),[t,t]);
> ec1:=grcomponent(R(dn,dn),[r,r]);
> ec2:=grcomponent(R(dn,dn),[theta,theta]);
> ec3:=grcomponent(R(dn,dn),[phi,phi]);
\end{verbatim}
obtaining the four Einstein equations of the problem. The rest of the Ricci tensor components are zero.

A simple inspection of the above obtained four equations reveals that only two of them are independent.  Also we can eliminate the second order derivative of the $\lambda(r)$ function between $ec0$ and $ec1$. These can be checked by using the next command lines
\begin{verbatim}
> expand(simplify(ecu2-ecu3/sin(theta)^2));
> ecu0;ecu1;ecu2;
> l2r:=solve(subs(diff(lambda(r),r,r)=l2r,ecu0),l2r);
> expand(simplify(subs(diff(lambda(r),r,r)=l2r,ecu0)));
> ecu11:=expand(simplify(subs(diff(lambda(r),r,r)=l2r,ecu1)));
\end{verbatim}
The first of the above commands checks the equality between $ecu2$ and $ecu3$, which simply gives a "0" (zero) and the next ones simply display the remaining three equations. The command that follows extracts the second order derivative of $\lambda(r)$ from $ecu0$  (substituting it with an intermediate constant $lr2$) and the next one substitutes the result in $ecu1$ and we obtained a  new
 equation $ecu11$. Thus we have now only two equations, $ecu11$ and $ecu2$, namely
\begin{equation}\label{ecu11}
\frac{2}{r}\partial_r\lambda(r)+\frac{2}{r}\partial_r\mu(r)=0
\end{equation}
\begin{equation}\label{ecu2}
1-e^{-2\mu(r)}+ \left[ r\,\partial_r\mu(r)-r\,\partial_r\lambda(r) \right]e^{-2\mu(r)}=0
\end{equation}
These two  differential equations need to be solved next, in order to obtain the solution. Of course we can now follow a classical strategy solving them manually as is done in any textbook (see \cite{8} for example). But it is also possible to continue with Maple, using the
$>dsolve$ command for solving differential equations (including systems of differential equations). Thus we write the next command
\begin{verbatim}
> dsolve({ecu11,ecu2},{mu(r),lambda(r)});
\end{verbatim}
which gives us the  function $\mu(r)$ as
\begin{eqnarray}
\mu(r)=\frac{1}{2} ln \left( \frac{r}{r e^{C1}
-1} \right) +\frac{C1}{2}
\nonumber
\end{eqnarray}
In particular a close inspection of both equations $(ecu11, ecu2)$ reveals that $ecu11$ is simply a relation between  the derivatives
of the two functions, namely
\begin{eqnarray}
\frac{d \lambda (r)}{d r} +\frac{d \mu (r)}{d r} =0 \nonumber
\end{eqnarray}
This shows that we will need only one integration constant and we can solve the equations only for one function as done above.
One can write the constant $C1$ as
\begin{eqnarray}
C1= ln \left( \frac{1}{r_s} \right) \nonumber
\end{eqnarray}
where the new introduced constant $r_s$ will be determined later. With these we can rewrite the $ecu2$ and applying again the $>dsolve$ command on it we obtain in the Schwarzschild (type) solution as
\begin{equation}\label{FSS}
ds^2 = \left(1-\frac{r_s}{r} \right)c^2dt^2 + \left(1-\frac{r_s}{r}\right)^{-1}dr^2 + r^2\left[d\theta^2 + sin^2(\theta)d\phi^2 \right]
\end{equation}
The constant $r_s$ is known under the name of Schwarzschild radius and can be determined using the newtonian limit of the field equations as it is done in any textbook (see \cite{8} for example). The precise value of it is $r_s= 2MG/c^2 $ but this has nothing to do with algebraic computing.

\section{Example 2: Simple cosmological models}

Modern cosmology is based on general relativity and Einstein equations, from which we can derive the so called Friedman equations. The latter ones form the core of all cosmological models. The most used metric for describing the dynamics of the universe in a cosmological model is the Friedman-Robertson-Walker metric (FRW), which in spherical coordinates has the following line element \cite{9}
\begin{equation}\label{cm1}
ds^2=c^2dt^2-a^2(t)\left[\frac{dr^2}{1-kr^2}+r^2(d\theta^2+\sin^2\theta d\phi^2)\right]
\end{equation}
where $k$ is the curvature constant and we are using the $(+,-,-,-)$ signature for the metric. Usually, this $k$ constant is taken to be $1$ (for closed universes), $-1$ (for open universe) and $0$
for a flat one. In (\ref{cm1}) we denoted by $a(t)$ the scale factor, that in the end will be the only unknown function of a cosmological model. This scale factor is directly related to the evolution of the universe. For the FRW metric (obtained form the cosmological principle i.e. the universe is spatially homogenous and isotropic) the scale factor is a function only on time. By introducing FRW metric (\ref{cm1}) into the Einstein equations (\ref{EE}) and assuming that the energy-momentum tensor is of a prefect fluid form \cite{8}
\begin{equation}\label{rm3}
T^{\mu\nu}=\left( \rho+\frac{p}{c^2} \right) u^\mu u^\nu-p g^{\mu\nu}
\end{equation}
one arrives to the Friedman-Lemaitre equations
\begin{equation}\label{rm4}
\begin{split}
&\ddot a=-\frac{4\pi G}{3}\left(\rho+\frac{p}{c^2} \right)a + \frac{1}{3}\Lambda c^2a\\
&\dot a^2=\frac{8\pi G}{3}\rho a^2+\frac{1}{3}\Lambda c^2a^2-c^2k
\end{split}
\end{equation}
If the cosmological constant $\Lambda$ is set to zero in eqs.(\ref{rm4}) then the equations are called simply the Friedman equations. In eq. (\ref{rm3}) $u^\mu$ represnets the 4-velocity of the cosmological fluid, while $p$ and $\rho$ stands for the pressure, respectively the mass density of the fluid.

In the same manner as done in \cite{3} we can compose a sequence of GRTensorII commands for obtaining the Friedman equations (\ref{rm4}). A student can write the program on a computer in less than an hour to the a job that if it is done by hand calculations it will take several good
hours to a very good student. The basic lines of the GRTensorII program are as follows:
\begin{verbatim}
> restart; > grtw(); qload(metrica_FRW);
> grdef(`u{^a}:=[1,0,0,0]`);
> grdef(`T{^a ^b}:=(rho(t)+p(t)/c^2)*u{^a}*u{^b}-p(t)*g{^a ^b}`);
> ...
> ec0:=R1_0-T1_0; > ec1:=R1_1-T1_1;
> ec1:=subs(diff(a(t),t,t)=-(1/6)*K*c^4*rho(t)*a(t)-
 (1/2)*K*c^2*p(t)*a(t)+(1/3)*Lambda*c^2*a(t),ec1): ec0; ec1;
\end{verbatim}
The first line of commands starts the GRTensorII and loads the FRW metric (\ref{cm1}). In the next lines we define the 4-velocity and the stress-energy tensor (\ref{rm3}) of the cosmological fluid. Now follows a series of more technical commands which can be found in the supplementary web material \cite{16}, commands that allows us to calculate and write the final form of Friedman equations (\ref{rm4}).

Friedman equations (\ref{rm4}) are in fact a system of two differential equations with three unknowns: $a(t)$, $p$ and $\rho$. Thus
one needs to find a third equation in order to completely solve the problem. In standard cosmology we use as a third relation the
equation of state
\begin{equation}\label{rm5}
p(t)=w\rho(t)c^2
\end{equation}
where $w$ is a constant ( $w=0$ for pressureless 'dust', $w=1/3$ for radiation and $w=-1$ for vacuum).

Let us further introduce the dimensionless quantities (see for example \cite{9}), usually called density parameters, which are defined by
\begin{equation}\label{rm5}
\Omega_i(t)\equiv \frac{8\pi G}{3H^2(t)}\rho_i(t)
\end{equation}
where $ H(t)=\dot a(t)/a(t)$ is known as the Hubble parameter and $i$ stands for matter, radiation and the cosmological constant
$\Lambda$. Besides these three quantities one can also define a curvature density parameter
\begin{equation}\label{rm5}
\Omega_k(t)= -\frac{c^2k}{H^2(t)a^2(t)}
\end{equation}
Rewriting the second equation of (\ref{rm4}) in terms of the new defined density parameters we arrive at a very simple expression

\begin{equation}\label{rm6}
\Omega_m + \Omega_r + \Omega_{\Lambda} + \Omega_k = 1
\end{equation}
Taking all the above into account and introducing a normalised scale factor $ A(t)=a(t)/a_0(t_{now})$ (with $a_0(t_{now})$
representing the value of the scale factor at the present epoch) we can finally write an equation for the evolution of the scale
factor, namely
\begin{equation}\label{rm7}
H^2(t)=H_0^2(\Omega_{r,0}a^{-4}+ \Omega_{m,0}a^{-3}+ \Omega_{k,0}a^{-2} +\Omega_{\Lambda,0})
\end{equation}
where $\Omega_{i,0}$ are the values of the densities measured at the present epoch.

Below we give the main GRTensorII command lines with the help of which we arrive at equation (\ref{rm7}). The complete sequence of commands can be found in the supplementary web material \cite{16}.
\begin{verbatim}
> ec1:=subs(diff(a(t),t)=H(t)*a(t),ec1);
> rho(t):=rho_m0*(a_0/a(t))^3+rho_r0*(a_0/a(t))^4;
> ec1:=subs(Omega_k0=1-Omega_m0-Omega_r0-Omega_Lambda0,ec1);
\end{verbatim}
We can now use the other numerical and computational facilities of Maple in order to numerically solve equation (\ref{rm7}) and
express the results as plots (see Fig. 1) of the scale factor as a function of cosmic time. For that we use the following code (see also the supplementary web material \cite{16}):
\begin{verbatim}
> ecu:=diff(A(t), t)-sqrt(Omega_m0/A(t)+Omega_r0/A(t)^2+
  Omega_Lambda0*A(t)^2+1-Omega_m0-Omega_r0-Omega_Lambda0);
> ecu_a:=subs(Omega_m0 =0.3,Omega_Lambda0=0.7,Omega_r0=0,ecu);
> sys1:={ecu_a,A(0)=1}: > f1:=dsolve(sys1,numeric):
> odeplot(f1,t=-2..2,axes=boxed,numpoints=1000,color=black);
\end{verbatim}

\begin{figure}[h!t]
    \centering
    \begin{subfigure}[b]{0.47\textwidth}
        \centering
             \includegraphics[width=\textwidth]{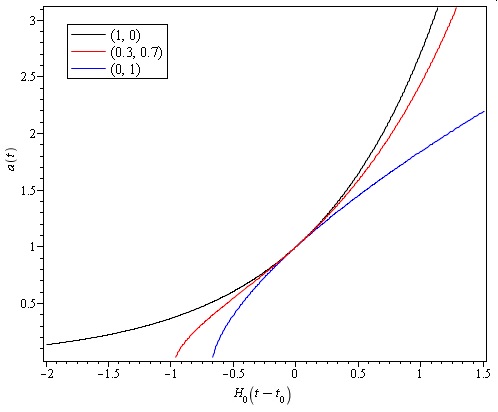}
              \caption{for different values of $(\Omega_{m,0},\Omega_{\Lambda,0})$ \\
               \ \  }
               \label{fig1}
        \end{subfigure}
        \quad
        \begin{subfigure}[b]{0.47\textwidth}
        \centering
                \includegraphics[width=\textwidth]{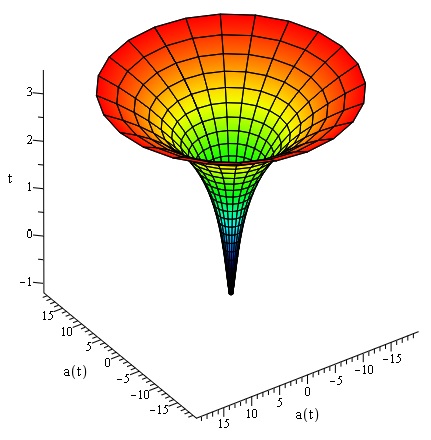}
                \caption{for ($\Omega_{m,0}=0.3,\Omega_{\Lambda,0}=0.7$) using cylindrical coordinates }
                \label{fig2}
        \end{subfigure}
\caption{Time evolution of the scale factor}
\label{fig1ab}
\end{figure}

\section{Conclusions and further developments}

The article describes a way in which some simple computer programs in Maple and GrTensorII can be used in teaching  GR and cosmology.
It is obvious that the speed of learning main concepts in GR (and subsequently differential geometry) can be successfully enhanced,
avoiding large hand computation steps and a lot of natural mistakes. On the  other hand it is clear that the lectures will need
to take place in a computer lab, which can be another way to increase the attractiveness of GR and cosmology.  We illustrated our experience with short and simple commands without  sophisticated tricks normally a professional in the field is using  (for example
building procedures and libraries via the symbolic computation facilities of Maple).

The small and short programs we described here can be also used as a strong basis for further developments in view of
more sophisticated and advanced examples. For instance one can develop the above procedures for cosmology in generalised
theories of gravity, like those with higher order Lagrangians (as we done in \cite{6}).

\section*{Acknowledgements}

This work was supported by a grant of the Romanian National Authority for Scientific Research, Programme for research-Space Technology and Advanced Research-STAR, project nr. 72/29.11.2013 between Romanian Space Agency and West University of Timisoara.

C.A.Sporea was supported by the strategic grant POSDRU/159/1.5/S/137750, Project “Doctoral and Postdoctoral programs support for increased competitiveness in Exact Sciences research” cofinanced by the European Social Found within the Sectorial Operational Program Human Resources Development 2007 – 2013.

\end{document}